\documentclass{kluwer}    
\usepackage{graphicx}

\begin{document}
\begin{article}
\begin{opening}
\title{From Large to Small Scales: Global Models of the ISM}
\author{Miguel \surname{de Avillez}}
\institute{Department of Mathematics, University of \'Evora, R. Rom\~ao Ramalho
59, 7000 Evora, Portugal.\email{mavillez@galaxy.lca.uevora.pt} }
\author{Dieter \surname{Breitschwerdt}}
\institute{Max-Planck-Institut f\"ur Extraterrestrische Physik,
        Giessenbachstra{\ss}e, Postfach 1312, 85741 Garching, Germany.\email{breitsch@mpe.mpg.de}}

\runningauthor{M. de Avillez \& D. Breitschwerdt} \runningtitle{Global Models of the ISM}
\date{November 30, 2002}

\begin{abstract}
  We review large scale modelling of the ISM with emphasis on the
  importance to include the disk-halo-disk duty cycle and to use a
  dynamical refinement of the grid (in regions where steep variations
  of density and pressure occur) for a realistic modelling of the ISM.
  We also discuss the necessity of convergence of the simulation
  results by comparing 0.625, 1.25 and 2.5 pc resolution simulations
  and show that a minimum grid resolution of 1.25 pc is required for
  quantitatively reliable results, as there is a rapid convergence for
  $\Delta x \leq 1.1$ pc.

\end{abstract}

\end{opening}

\section{Introduction}
\label{intro} The interstellar gas exhibits time-dependent
structures on all scales and is heavily non-linear and subject to
numerous instabilities in certain parameter regimes, which may lead to
chaotic behaviour, while in others the evolution is fairly
predictable. The key to a realistic description is the highest
possible spatial resolution and a realistic input of the basic
physical processes with appropriate boundary conditions. The former is
required due to the formation of small scale structures resulting from
instabilities in the flows, in particular from thermal instabilities
and condensations as a result of radiative cooling. The amount of
cooling may be substantially increased if the resolution is high
enough to trace regions of high compression by shocks, rather than
smearing them out over a larger volume thus decreasing the average
density. Such a realistic description requires therefore the
appropriate tools, which are computer clusters with parallelized HD
and MHD codes and a sophisticated method of tracking non-linear
structures, such as shock waves on the smallest possible scales in
conjunction with adaptive mesh refinement. Thus the global evolution
of the ISM resulting from mass, momentum and energy input due to
supernovae can be adequately described.

The structure of the paper is as follows. Section 2 deals with the
importance of the disk-halo-disk cycle for global modelling of the
ISM. In Section 3 the signature of the initial evolution in the
temperature and density PDFs is discussed, as well as the
resolution effects in the time histories of the volume filling
factors and the maximum density and minimum temperature in the
ISM. Moreover the necessity for convergence of the simulations is
addressed. In section 4 a summary and final remarks are given.

\section{Duty Cycle and Grid Sizes}

\subsection{The Duty Cycle}

The evolution of the ISM in disk galaxies is intimately related to the
vertical structure of the thick gas disk and to the input of energy by
supernovae per unit area in the Galactic disk. The system can evolve
towards some sort of dynamical equilibrium state on the global scale
if the boundary conditions vary only in a secular fashion. Such an
equilibrium is determined by the input of energy into the ISM by SNe,
diffuse heating, the energy lost by radiative cooling and magnetic
energy, and is only possible \emph{after} the full establishment of
the duty cycle of the warm and hot gas and its circulation between the
disk and the halo, which takes several hundred Myr. An upper limit for
this can be estimated by calculating the flow time, $t_f$, that the
gas needs to travel to the critical point of the flow in a
steady-state (see Kahn 1981). This is the characteristic distance from
which information in a thermally driven flow can be communicated back
to the sources. Then $t_f \sim r_c/c_s$, where $r_c$ and $c_s$ are the
location of the critical point and the speed of sound, respectively.
For spherical geometry, the critical point can be simply obtained from
Bernoulli's or the steady state fluid equations, $r_c \sim G
M_{gal}/(2 c_s^2)$, which yields for a Milky Way mass of $M_{gal} = 2
\times 10^{11} \, {\rm M_{\odot}}$ a distance $r_c \approx 21.6$ kpc,
and thus $t_f \sim 1.5 \times 10^8$ years as an upper limit for the
ascending flow time. This value is of the same order as the radiative
cooling time, ensuring that the flow will not only cool by adiabatic
expansion, but also radiatively, thus giving rise to the fountain
return flow, which is the part of the outflow that loses pressure
support from below and therefore cannot escape from the galaxy. Note,
that $r_c$ is the maximum extension of the fountain.

In the disk, on the other hand, the heating and cooling time scales
are much shorter than in the halo, since they scale typically with the
gas density there, i.e. $\propto 1/n$ and $\propto 1/n^2$,
respectively.  Avillez \& Breitschwerdt (2003) show from large-scale
simulations of the ISM that typical time scales of pressure
fluctuations about a mean value of $P/k_B \sim 2600 \, {\rm cm}^{-3}
\, {\rm K}$ are about $30$ Myr, which is by the way a typical time
scale for superbubble evolution.

It should be emphasized, since disk and halo are coupled
dynamically, not only by the escape of hot gas, but also by the
fountain return flow striking the disk, that the disk equilibrium
will also suffer secular variations on time scales of the
order of 100 Myr.

\subsection{Vertical Extension of Computational Domain}

The need of a duty cycle and the establishment of a global dynamical
equilibrium requires the use of an extended grid in the direction
perpendicular to the Galactic plane. The lack of such an extended
$z$-grid inhibits the disk-halo-disk circulation of matter, which
otherwise would return gas to the disk sometime later, with noticeable
effects for the dynamical evolution there. In runs in which the
vertical extension is small, e.g. 1 kpc above and below the
midplane\footnote{Note, that the maximum extension of the fountain is
  more than an order of magnitude larger, and therefore such
  restricted calculations definitely miss an important component of
  the galactic gas dynamics.}, compared to the maximum height to which
the hot plasma rises due to the injection of energy and momentum from
the sources, most of the gas escapes from the disk in less than 100
Myr (without ever returning to it), unless the loss of matter is
compensated by some injection of mass into the grid through the top
and bottom boundaries. This can be accomplished by the use of (1)
periodic boundary conditions (the gas leaving the top/bottom boundary
enters in the bottom/top boundary), (2) reflection boundary conditions
where the gas hitting the top/bottom boundary reflects in it returning
back, or by (3) calculating the amount of mass leaving the grid on each
boundary and inject that mass in the form of clouds into the
top/bottom boundaries. These schemes assume that the amount of gas
that leaves the grid is the same that enters it, neglecting that some
of this matter may actually escape from the Galaxy altogether as a
wind\footnote{In the case in which the appropriate grid is used there
  is still gas leaving the upper and bottom boundaries amounting to
  some 10\% of the total mass with which the simulations started
  (Rosen \& Bregman 1995, Avillez 2000). Thus, a smaller grid with
  open boundaries on the top and bottom of the grid will imply the
  loss of a \emph{large amount} of matter within a short time thereby
  evacuating the gas disk.} or return to the disk at a different
locations due to colliding gas streams. The wind effect may be
included in scheme (3) by assuming that the gas injected into the grid
is only a fraction of the escaping gas mass.

In any case these schemes neglect the effects of the duty cycle in
the dynamics of the disk gas as there is no recycling of gas, and
as a consequence the simulations can only run for a small
evolution time thus reflecting the \emph{initial evolution} of the
system. Therefore, the volume weighted histograms of the
thermodynamic properties (e.g., density, pressure and temperature)
of the disk gas will retain a memory of the original set-up.

\subsection{Grid Extension Parallel to the Midplane}

The extension of the grid in the direction parallel to the Galactic
plane is also an important factor. Large $xy$-grids, parallel to the
midplane are required because counter pressure effects are very
important in the bubbles confinement (Avillez 2000; Avillez \&
Breitschwerdt 2003). If the grid in the plane is too small, say less
than 500 pc, then large scale horizontal flows resulting from
superbubbles expanding into low density regions and eventually
colliding with other hot tenuous bubbles will not be captured
adequately. We therefore consider an extension of 1 kpc $\times$ 1 kpc
as the minimum grid extension in the plane.

\section{Global Modelling of the ISM}

\subsection{SN-Driven ISM Model}

In order to gain insight into the nonlinear physics of a complex
system like the ISM, driven by temporally and spatially variable
energy input we first try to understand the interplay and effects
of the most dominant processes, and therefore only the most
important ingredients are included in the models, neglecting other
processes that may be relevant too. We therefore, do not strive
for completeness, but for a better understanding of the ISM gas
dynamics. The discussion that follows is based on the results from
three-dimensional kpc-scale modelling of the ISM with a grid
centred on the solar circle and extending from $z=-10$ to 10 kpc
with a disk area of 1 kpc$^{2}$. These simulations use adaptive
mesh refinement with resolutions of 2.5, 1.25 and 0.625 pc and
were obtained with a modified version of the 3D SN-driven ISM
model of Avillez (2000).

The basic processes included in the model are the gravitational
field provided by the stars in the disk, radiative cooling
(assuming an optically thin gas in collisional ionization
equilibrium) with a temperature cut off of 10 K, and uniform
heating due to starlight varying with $z$. In the Galactic plane
background heating is chosen to initially balance radiative
cooling at 9000 K. With the inclusion of background heating the
gas at $T<10^{4}$ K becomes thermally bistable. The interstellar
gas is initially set up with a density stratification distribution
that includes the cool, warm, ionized and hot gases according to
observations. The prime sources of mass, momentum and energy are
supernovae types Ia, Ib+c and II with scale heights, distribution
and rates according to observations. OB associations can be setup
in regions with density and temperature thresholds of 10 cm$^{-3}$
and 100 K, respectively, and an initial mass function is applied
to determine the number of OB stars and their masses, forming an
OB association. The time interval between the explosions of all OB
stars is determined by their main sequence life time, and the
kinematics of both the association and of the low mass stars in
the association can be followed in detail.

\subsection{General Evolution}

Our simulations reproduce many of the features that have been observed
in the Milky Way and other star forming galaxies, namely: (i) A thick
frothy gas disk composed of a warm, neutral medium overlying a thin HI
disk, with a variable thickness up to $\sim 80$ pc; (ii) bubbles and
superbubbles and their shells distributed on either side of the
midplane; (iii) tunnel-like structures (chimneys) crossing the thick
gas disk and connecting superbubbles to the upper parts of the thick
gas disk; (iv) thick gas disk with a distribution compatible with the
presence of two phases having different scale heights: a neutral layer
with $z_n \sim 500$ pc (warm HI disk) and an ionized component
extending to a height $z_i \sim 1.5$ kpc above the thin HI disk; (v)
cold gas mainly concentrated into filamentary structures running
perpendicular to the midplane. These clouds form and dissipate within
some 10-12 Myr. Compression is the dominant process for their
formation, but thermal instability also plays a r\^ole; (vi) for
supernova rates $\sigma/\sigma_{Gal}\leq 2$, $\sigma_{Gal}$ is the Galactic
value, the hot gas has a moderately low volume filling factor in
agreement with observations ($\sim 20\% $) even in the absence of
magnetic fields and is mainly distributed in an interconnected tunnel
network and in some case it is confined to isolated bubbles (see
Figure~\ref{mavillez_fig8}).

\subsection{Effects of the Initial Evolution}

\begin{figure}[t]
\centering
\includegraphics[width=0.5\hsize,angle=0]{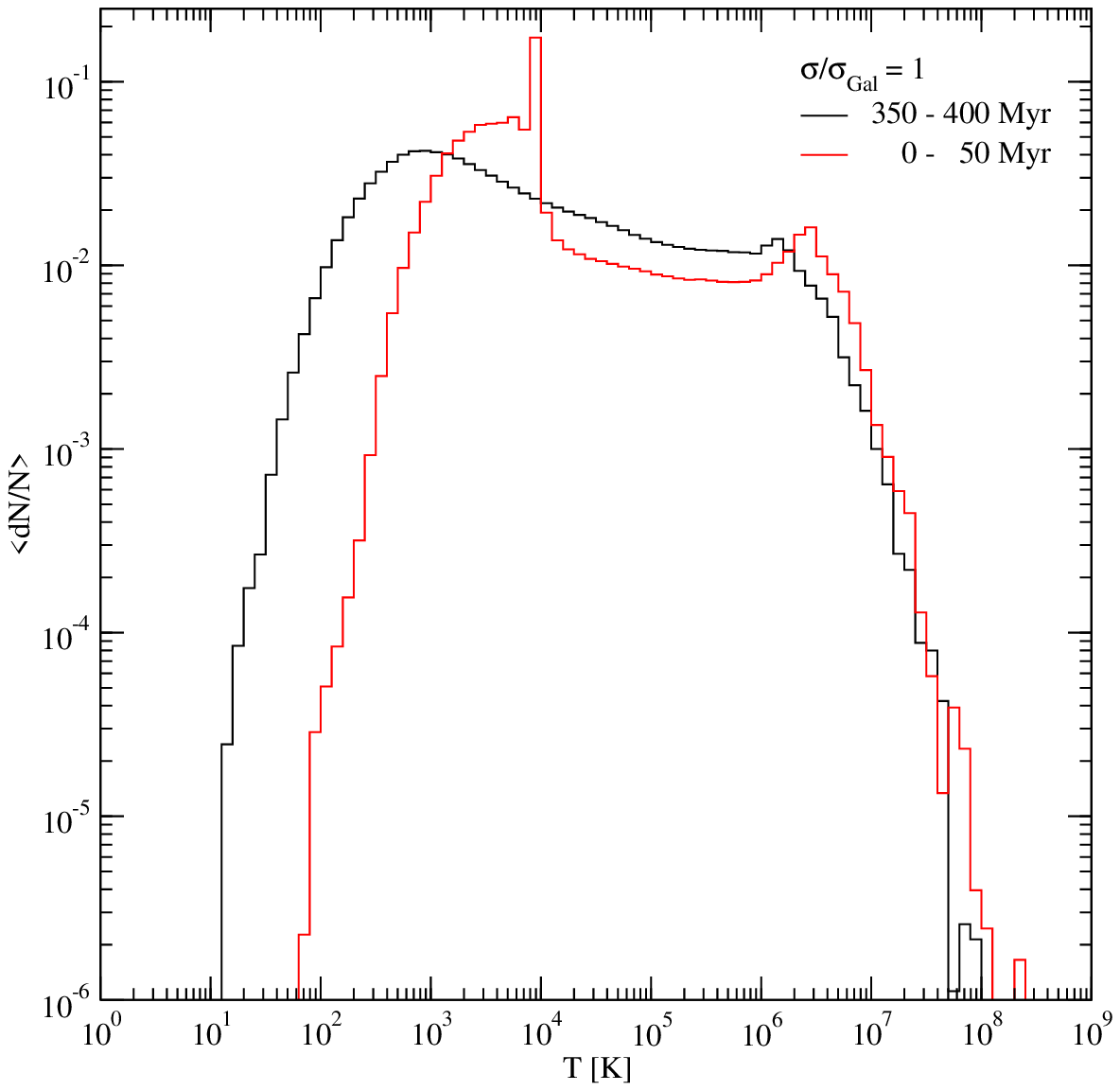}\includegraphics[width=0.5\hsize,angle=0]{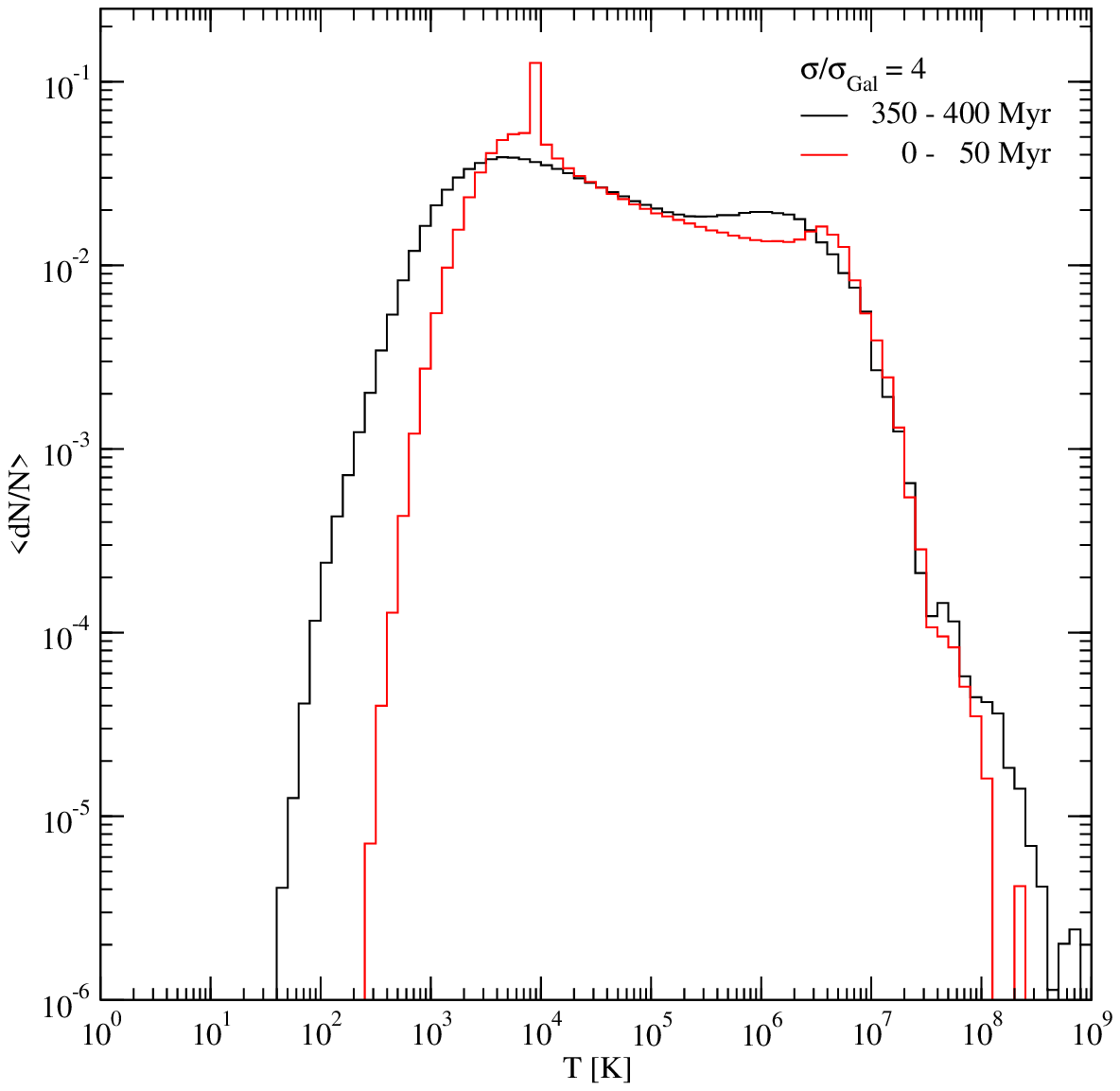}
\caption{Averaged volume-weighted temperature PDFs over the
periods of 0-50 Myr (red) and 350-400 Myr (black) calculated using
51 snapshots taken at time intervals 1 Myr. The supernova rates
used in these models are: $\sigma/\sigma_{Gal}=1$ (left panel) and
4 (right panel). The resolution of the finest AMR level is 1.25
pc.} \label{temppdfs-1.25pc}
\end{figure}

A major consequence of the set-up of the duty cycle is that the system
loses any memory of its initial conditions and evolution as most of
the disk gas has already travelled into the halo and back to the disk.
Thus, as should be in representative global simulations, memory
effects are not present in the volume weighted histograms (PDFs) of
the different thermodynamic properties as well as in the history of
the volume filling factors of the different ISM phases. Figure
\ref{temppdfs-1.25pc} compares PDFs of the temperature over the
periods of 0-50 Myr (red) and 350-400 Myr (black) for two AMR
simulations of the ISM using supernova rates $\sigma/\sigma_{Gal}=1$
(left panel) and 4 (right panel). The finest AMR level resolution is
1.25 pc.

These PDFs indicate that for low $\sigma$, the temperature peak is at
about 2000 K, making the cold/warm HI gas the most abundant gas phase,
consistent with observations.  The relative importance of the hot
phase increases with SN rate, and one is moving towards a bimodal
distribution. It should be emphasized that a simulation time of only
50 Myrs is not sufficient to establish this result, mainly because the
effect of upwards transport has as a duty cycle of the order of a few
hundred Myrs. In any of the cases shown in the figures the averaged
PDFs for the initial 50 Myr have two pronounced peaks, one around 8000
K and the other around $5\times 10^{6}$ K. Note that with the increase
of SN rate the PDFs of the first 50 Myrs suffer large variations,
indicating that the loss of recollection of the initial conditions
will occur earlier for the highest SN rates.

The inclusion of snapshots taken during the first tens of Myrs in the
calculation of PDFs, and consequently also for the volume filling
factors, has a considerable effect in the overestimation of the
coolest gas in the beginning of the simulations when cooling and
gravity are the main physical dominant processes as the system is only
\emph{building up} the pressure support against gravity. Thus, during
the first Myrs much of the gas collapses into the plane due to
gravity, and simultaneously cools towards low temperatures. As soon as
the pressure has built up, there is a redistribution of matter in the
grid filling it. Therefore, the PDFs of the disk gas including the
first say 40-50 Myr snapshots will show the presence of a large fraction of
cold gas, while the PDFs constructed at later times, say up to 90 Myr
or so, will show the dominance of the hot gas.  A combination of these
PDFs gives a pronounced bimodal distribution in the averaged PDFs
corresponding to large volume filling factors for the hot and coolest
ISM phases.  Moreover, the averaged PDFs of the temperature will show
a bimodal structure even if there is no background heating due to
starlight. However, it should be emphasized that global ISM models,
which may be considered representative for star forming galaxies,
should be free of initial set-up features.

\subsection{Resolution and Convergence}
\begin{figure*}[thbp]
\centering
\vspace*{1.1in}
Left panel: mavillez$\_$fig2a, Right panel: mavillez$\_$fig2b
\vspace*{1.1in}
\caption{Density distribution in the Galactic plane for AMR
resolutions of 2.5 (left panel) and 0.625 pc (right panel) at
times 400 and 384 Myr of disk evolution, respectively. The SN rate
is twice the Galactic value. High density gas is shown in blue, while low density (high temperatures) is shown in red. For contrast purposes, the maximum density cutoff is set at $10^{2.2}$ cm$^{-3}$.} \label{mavillez_fig8}
\end{figure*}

The simulations show how crucial spatial resolution is in order to
capture small scale structures and, in particular, the cold gas as can
be seen by the comparison between the left and right panels of
Figure~\ref{mavillez_fig8}. Both panels show a snapshot of the density
distribution in the Galactic plane for $\sigma/\sigma_{Gal}=2$, at
different evolution times, for runs with AMR resolutions of 2.5 pc
(left panel) and 0.625 pc (right panel). High density gas is shown in
blue, while low density gas (high temperatures) is shown in red.  For
contrast purposes, the maximum density cutoff is set at $10^{2.2}$
cm$^{-3}$, although in the 0.625 pc resolution image there are regions
where the density is as large as 800 cm$^{-3}$. The panels show a
clear difference between the morphology of the ISM in the two
resolutions. In the 0.625 pc resolution image the filamentary
structures are well defined and thinner than in the left panel. It can
be clearly seen how cooling promotes the formation of small scale
structures.

\begin{figure*}
\centering
\includegraphics[width=0.51\hsize,angle=0]{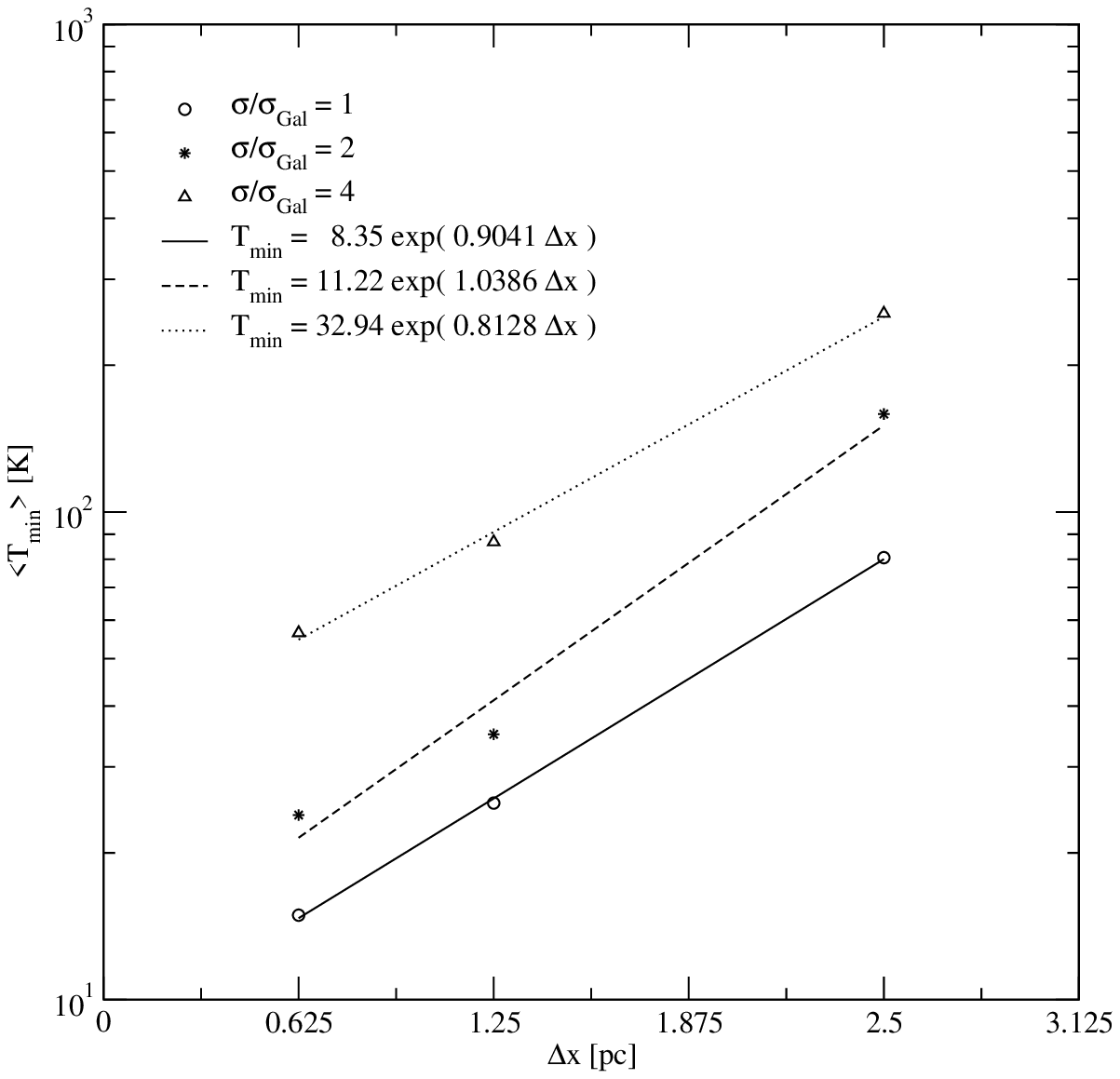}\includegraphics[width=0.51\hsize,angle=0]{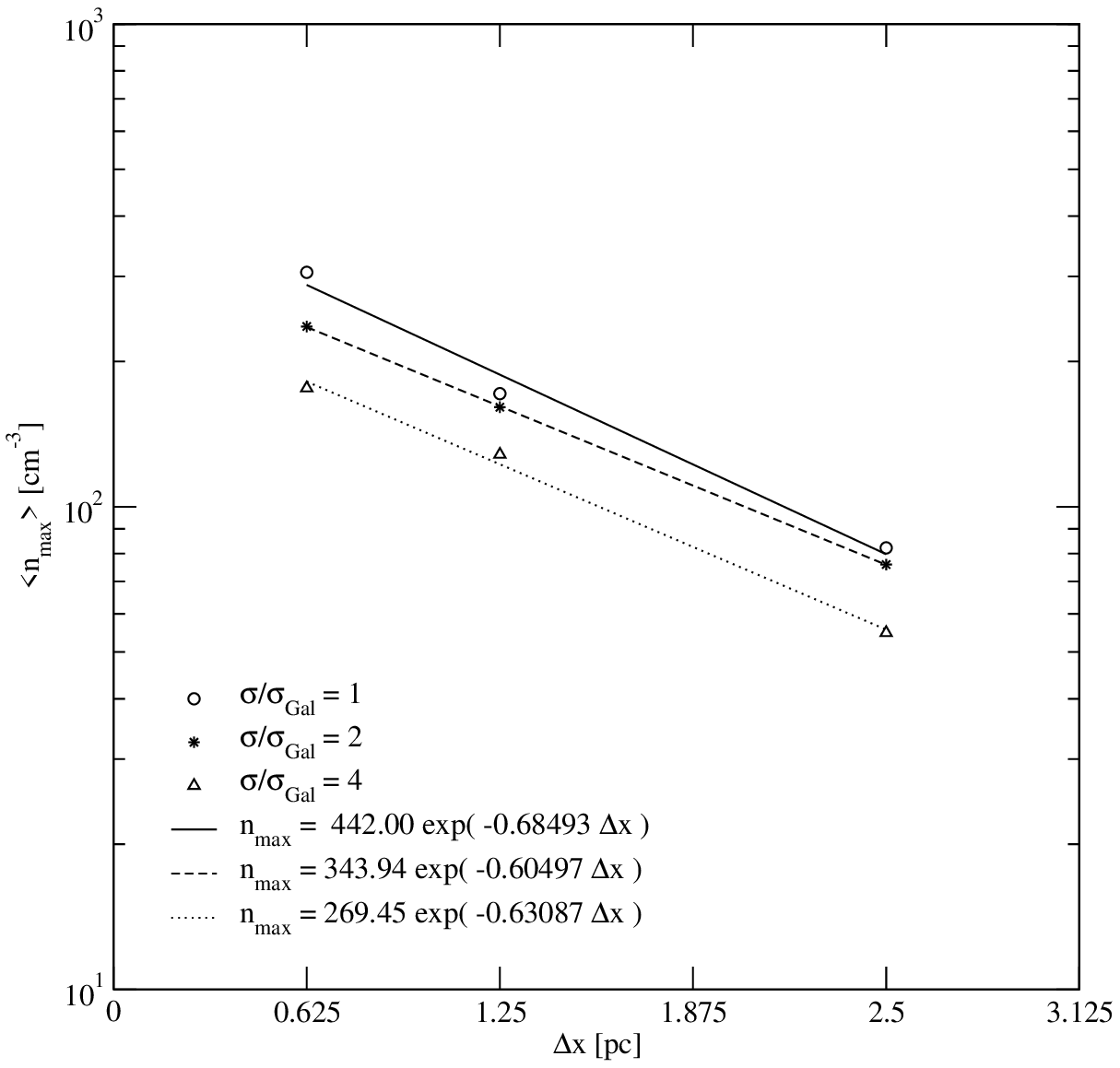}
\caption{Comparison between the average minimum temperature (left panel) and
  maximum density (right panel) as function of resolution for the
  three SN rates: $\sigma/\sigma_{Gal}=1$ (open circles),
  $\sigma/\sigma_{Gal}=2$ (stars), and $\sigma/\sigma_{Gal}=4$
  (triangles). The plots also show exponential fits to the data
  points. These average values were calculated during the last 200 Myr
  of evolution, such that their history does not reflect any
  recollection of the initial conditions, and after the establishment
  of the duty cycle.}
\label{resolution}
\end{figure*}

A comparison between the maximum density and minimum temperature
measured at the finer level resolutions of 0.625, 1.25 and 2.5 pc
reveals that an increase in resolution by a factor of two from $\Delta
x=2.5$ to 1.25 pc implies an increase in the maximum density and a
decrease in minimum temperature of the gas by factors greater than 4.
With an increase in resolution from 1.25 to 0.625 pc the decrease in
$T_{min}$ and increase in $n_{max}$ is smaller than a factor 2 as can
be seen in Figure ~\ref{resolution}. The figure compares the average
values of $T_{min}$ and $n_{max}$ calculated between 200 and 400 Myr
of evolution for the three resolutions. The values measured for
$\sigma/\sigma_{Gal}=1$ are shown by open circles, for
$\sigma/\sigma_{Gal}=2$ by stars, and for $\sigma/\sigma_{Gal}=4$ by
triangles. The straight lines show the exponential fits associated to
the variation of $T_{min}$ (decrease with resolution) and $n_{max}$
(increase with resolution).

Resolution also affects the occupation fraction of the gas that is
observed in the different ISM phases, most importantly for the cold
and hot phases. Figure~\ref{vff-compare-01} compares the volume
filling factors of the different phases for the AMR resolution level
of 0.625 pc (black), 1.25 pc (red) and 2.5 pc (green) for
$\sigma/\sigma_{Gal}=2$ and 4. In general the increase of resolution
from 2.5 to 1.25 pc affects three of the ISM phases, and most
importantly affects the amount of cold gas (which increases from $9\%$ when 2.5 pc resolution is used to some $18\%$ for 1.25 pc or 0.625 pc resolutions, respectively) for
$\sigma/\sigma_{Gal}\leq 2$ after several crossing times. The relative
variation of the cold gas is still very high for
$\sigma/\sigma_{Gal}\geq 4$, whereas for the hot phase there is a
reduction of only a few percent of the occupation fraction with
resolution, since here compression and cooling are less effective.

The variations in the volume filling factors are easily understood if
one takes into account that with an increase of resolution it becomes
possible to resolve the smallest scale structures. The rate
of cooling may be substantially increased if the resolution is
high enough to allow for the development of turbulent shear layers in
addition to more contact surfaces. The latter is most important,
because it allows for a faster mixing between parcels of gas with
different temperatures (conduction or diffusion processes being
  of second order and hence inherently slow in nature). The mixing
here is promoted by numerical diffusion rather than molecular
diffusion, and therefore, the time scale for mixing to occur is
somewhat smaller (because it happens on larger scales) than
those predicted by molecular diffusion theory (e.g., Avillez \& Mac
Low 2002). However, turbulent diffusion will be most efficient.

\begin{figure*}
\centering
\vspace*{1.1in}
Left panel: mavillez$\_$fig4a, Right panel: mavillez$\_$fig4b
\vspace*{1.1in}
\caption{Comparison between volume filling factors of cold ($T<10^{3}$ K), cool
($10^{3}<T<10^{4}$ K), warm ($10^{4}<T<10^{5.5}$ K) and hot ($T>10^{5.5}$ K gas
for the finest AMR level grid resolutions of 0.625 pc (black), 1.25 pc (red),
and 2.5 pc (green) for the SN rates $\sigma/\sigma_{Gal}=2$ (top panels),
and 4 (bottom panels).}
\label{vff-compare-01}
\end{figure*}

The fact that there is only a small difference between the volume
filling factors for the 1.25 and 0.625 pc resolutions in addition to
the small variation of $T_{min}$ and $n_{max}$ is a clear indication
for the convergence of the results, reproducing the physical processes
involved in the dynamics and evolution of the ISM. Thus, the
simulations with a resolution of 1.25 pc (or higher) can represent the
real ISM. From Fig.~\ref{resolution} and the fit for $\langle T_{min}
\rangle$ and $\langle n_{max} \rangle$, e.g.\ $\langle T_{min} \rangle
= 8.35 \exp({\Delta x \over 1.106 {\rm pc}})$, we can see that there
is rapid convergence for $\Delta x \leq 1.1$ pc.

\section{Summary and Final Remarks}

In this paper we discus the effects of the inclusion of the duty
cycle in global models of the ISM and show that sufficient
resolution is crucial in order to obtain quantitatively reliable
results that can be compared to observations, and that a minimum
grid resolution of 1.25 pc is needed. This is unambiguously
demonstrated by the rapid convergence towards the 0.625 pc
resolution simulations. The occupation fraction of the different
ISM phases depends also sensitively on the presence of a duty
cycle established between the disk and halo acting as a pressure
release valve for the hot phase.

The calculations presented in this paper do neither include the
magnetic field nor the cosmic rays. A parameter study of their
effects on the ISM is underway and will be described in
forthcoming papers. If the magnetic field is present and is
initially mainly orientated parallel to the disk, transport into
the halo may be inhibited, although not prevented. As a
consequence the hot gas in the disk should have a slightly higher volume
filling factor than in the present simulations, since this is the
ISM component that tries to escape first from the disk. However,
on larger scales magnetic tension forces become weaker than on the
smallest scales and therefore vertical expansion might still take
place efficiently and therefore the occupation fraction of the hot
gas could still be comparable to the values observed in the
present simulations.

\end{article}
\end{document}